\documentclass[12pt]{article}
\pdfoutput=1
\usepackage[utf8]{inputenc}
\usepackage[DIV=13]{typearea}
\usepackage{ragged2e}
\usepackage{amsmath,amsfonts,bbm,mathrsfs,amssymb}
\usepackage{slashed,cancel}
\usepackage[usenames,dvipsnames]{xcolor}
\usepackage{cite}
\usepackage{bm} % boldphase en mathmode
\usepackage{hyperref}
\hypersetup{colorlinks=true,urlcolor=blue,anchorcolor=blue,citecolor=blue,filecolor=blue,
            linkcolor=blue,menucolor=blue, linktocpage=true,pdfproducer=medialab}
\usepackage[english]{babel}
\usepackage{catchfile}
\usepackage{indentfirst}
\usepackage[scriptsize]{subfigure}
\usepackage{graphicx}
\usepackage{float}
\usepackage{enumerate}
\usepackage{multirow}
\usepackage{tikz-feynman}
\usepackage[normalem]{ulem} 

\numberwithin{equation}{section}

\newcommand{\be}{\begin{equation}}
\newcommand{\ee}{\end{equation}}
\newcommand{\ba} {\begin{equation}\begin{aligned}}
\newcommand{\ea} {\end{aligned}\end{equation}}
\newcommand{\bg} {\begin{equation}\begin{gathered}}
\newcommand{\eg} {\end{gathered}\end{equation}}

\newcommand{\sL}{\mathscr{L}}

\newcommand{\cN}{\mathcal{N}}

\newcommand{\TeV}{\ \text{TeV}}
\newcommand{\GeV}{\ \text{GeV}}

\def\Tr{{\rm Tr}}

\def\UPQ{U(1)_\text{PQ}}

\begin{document}
	{\hfill CERN-TH-2022-182}

\vspace{2cm}

\begin{center}
	\boldmath
	
	{\textbf{\LARGE Axion-like ALPs}}
	
	\unboldmath
	
	\bigskip
	
	\vspace{0.5 truecm}
	
	{\bf Fernando Arias-Arag\'on}$^{a}$, {\bf J\'er\'emie Quevillon}$^{a,b}$, and {\bf Christopher Smith}$^{a}$
	\\[5mm]
	{$^a$\it Laboratoire de Physique Subatomique et de Cosmologie,}\\
	{\it Universit\'{e} Grenoble-Alpes, CNRS/IN2P3, Grenoble INP, 38000 Grenoble, France}\\[2mm]
	{$^b$\it CERN, Theoretical Physics Department, Geneva, Switzerland}\\[2mm]
	
	\vspace{2cm}
	
	{\bf Abstract }
\end{center}

\begin{quote}
The Effective Field Theory (EFT) Lagrangian for a generic axion-like particle (ALP) has many free parameters, and leaves quite some freedom for the expected phenomenology. In this work, we set up more constrained EFTs by enforcing true axion-like properties for the ALP. Indeed, though the Peccei-Quinn symmetry of the QCD axion is anomalous, it is so in specific ways, and this shows up as consistency conditions between the gauge boson and fermion couplings. We propose to enforce such conditions, as inspired from the DFSZ and KSVZ scenarios, on the generic ALP EFTs. These truly axion-like ALP EFTs are then particularly well-suited as benchmark scenarios, to be used in the search for ALPs both at colliders and at low-energy experiments.

\end{quote}

\thispagestyle{empty}
\vfill

\newpage

\tableofcontents

%%%%%%%%%%%%%%%%%%%%%%%%%%%%%%%%%%%%%%%%%%%%%%%%%%%%%%%%%%%%
\section{Introduction}
%%%%%%%%%%%%%%%%%%%%%%%%%%%%%%%%%%%%%%%%%%%%%%%%%%%%%%%%%%%%
\label{sec:Intro}

Despite its astonishing predictive power, the Standard Model (SM) still presents some issues that require the inclusion of new physics. One of such open problems is the so-called Strong CP Problem, which stems from the non-observation of CP violation in the strong interactions in the form of an electric dipole moment for the neutron~\cite{Abel:2020pzs}.

The Peccei-Quinn (PQ) mechanism~\cite{Peccei:1977hh} considers the inclusion of a new axial global symmetry, $U(1)_{PQ}$. Together with its associated Goldstone boson, the axion $a$~\cite{Wilczek:1977pj,Weinberg:1977ma}, they represent arguably one of the most appealing solutions to this problem. In recent years, the interest in axions has grown and extended itself to axion-like particles, or ALPs. These new hypothetical particles share the basic features of axions, being pseudoscalars with dimension-five couplings to gauge bosons, but as a consequence of some additional explicit symmetry breaking term, they are free from the relation between mass and symmetry breaking scale that is predicted for the QCD axions.

Currently, an increasing experimental task force is being born aiming to set bounds on the several couplings of ALPs to both fermions and gauge bosons of the SM. Additionally, these particles and their couplings have been extensively studied in the literature~\cite{Mimasu:2014nea,Bauer:2017ris,Brivio:2017ije,Alonso-Alvarez:2018irt,Bauer:2018uxu,Gavela:2019cmq,Gavela:2019wzg,Alves:2019xpc,Harland-Lang:2019zur,Chala:2020wvs,Bauer:2020jbp,Bauer:2021mvw,Galda:2021hbr,Brivio:2021fog,Bonilla:2021ufe,Bonilla:2022pxu,Bonnefoy:2022rik,Kling:2022ehv,Bharucha:2022lty}, following an Effective Field Theory (EFT) approach. Our goal in this paper is to analyze this formulation, and identify promising benchmarks.

Typically, ALP EFTs are constructed by adding to the SM the kinetic terms for a massive neutral scalar, and two sets of dimension-five couplings:
\ba
\sL_{ALP} =& \frac{1}{2}\left(\partial_\mu a\partial^\mu a-m_a^2 aa\right)-i\displaystyle\sum_{f} \frac{\chi_f}{v_a} \partial^\mu a \bar{f}\gamma_{\mu} f
\\
&+ \frac{a}{16\pi^2 v_a}\left(g_s^2\cN_C G_{\mu\nu}^a\Tilde{G}^{a,\mu\nu}+g^2\cN_L W_{\mu\nu}^i\Tilde{W}^{i,\mu\nu}+{g^\prime}^2\cN_Y B_{\mu\nu}\Tilde{B}^{\mu\nu}\right),
\label{eq:ALPEFT}
\ea
where the sum runs over all the SM chiral fermions,  $\tilde{X}^{\mu\nu}=\frac{1}{2}\varepsilon^{\mu\nu\alpha\beta}X_{\alpha\beta}$ for $X=G,B,W$, $\chi_f$ and $\cN_{C,L,Y}$ are unknown Wilson coefficients, $v_a$ is the ALP decay constant, typically much larger than the ALP mass $m_a$. The couplings $\chi_f$ could, in principle, be matrices in flavour space, but for this work we will take them proportional to the identity matrix, restricting ourselves to flavour conserving and universal scenarios. For this construction, one assumes the SM gauge symmetries to remain exact, and constrains the ALP couplings to exhibit only axion-like breaking of the shift symmetry $a \rightarrow a + v_a \theta$ for any constant $\theta$. In other words, the ALP gauge couplings mimic the so-called axion anomalous couplings to the gauge field strengths, while those to the SM fermions have to be shift-symmetric, thus involving the derivative of the axion field. Once such an EFT is prescribed at the EW scale, it is then run down to the $m_a$ scale relevant for light ALP searches. At one loop, the gauge and fermion couplings to an ALP mix via typical triangle graphs, allowing one to set limits on all the couplings out of the bounds for any of the ALP decay channel. For such analyses, see Refs.~\cite{Bauer:2017ris,Bauer:2021mvw,Bonilla:2021ufe}. 
The generic EFT Lagrangian in Eq.~\eqref{eq:ALPEFT} has many free parameters, and leaves quite some freedom for the expected ALP phenomenology. In the present paper, our goal is to set up more constrained EFTs, to be used as benchmark scenarios to search for ALPs. Our guiding principle is to enforce true axion-like properties for the ALP. Indeed, though the QCD axion PQ symmetry is anomalous, it is so in specific ways, and this shows up as consistency conditions between the gauge boson and fermion couplings. Its EFTs are not as generic as that in Eq.~\eqref{eq:ALPEFT}. Specifically, in the Dine, Fischler, Srednicki, Zhitnitsky (DFSZ)~\cite{Dine:1981rt,Zhitnitsky:1980tq} scenario, the axion couplings are anomaly-free, in the sense that the axion couples to ordinary matter only through pseudoscalar couplings to the massive SM fermions. By contrast, in the Kim, Shifman, Vainshtein, Zakharov (KSVZ)~\cite{Kim:1979if,Shifman:1979if} case, the axion inherits only anomalous gauge couplings from its coupling to some new heavy coloured fermion at one loop, while those to SM fermions arise only at two loops, via anomaly-free finite diagrams. These are the constraints we want to impose to construct DFSZ and KSVZ ALP scenarios. Alleviating these fundamental assumptions, one may face Ultra-Violet models suffering from serious inconsistencies.

The structure of this letter is the following: in Sec.~\ref{sec:DFSZ} we will describe the setup for a generic DFSZ-like ALP scenario, followed by a dedicated discussion of a more constrained scenario reminiscent of the DFSZ Axion. In each case, we will investigate the true number of free parameters present in the theory as well as the possible correlations among loop-induced couplings. Sec.~\ref{sec:KSVZ} will follow the same procedure for KSVZ ALPs, starting with a generic description of its Lagrangian, followed by a detailed discussion of the loop generated couplings. Finally, Sec.~\ref{sec:Conclusions} closes our analysis with some concluding remarks.

%%%%%%%%%%%%%%%%%%%%%%%%%%%%%%%%%%%%%%%%%%%%%%%%%%%%%%%%%%%%
\section    {DFSZ-like ALPs}\label{sec:DFSZ}

In the original DFSZ scenario, the axion couplings to SM gauge bosons and fermions arise entirely from new couplings of a complex scalar field $\phi_{PQ}$ to the two Higgs doublets present in that framework. In practice, this means that at tree level, the axion shows up only in the Yukawa couplings of the SM fermions. This can be summarized by noting that the DFSZ axion field simply receives a tiny component aligned with the pseudoscalar Higgs field of the usual two-Higgs doublet model (2HDM), $A$. Indeed, provided the PQ symmetry is active, $A$ is massless, and mixes with the pseudoscalar component of $\phi_{PQ}$. Importantly, though couplings like $aaW_\mu^+ W_\nu^-g^{\mu\nu}$ or $aaZ_\mu Z_\nu g^{\mu\nu}$ arise from the kinetic term of the Higgs doublets, couplings of a single axion to two gauge bosons do not appear at tree level, which are the ones we will be interested in during the following section.

\subsection{Generic Setup}\label{sec:DFSZ-Setup}

Let us now discuss the generic case of an ALP arising in a scenario where $U(1)_{PQ}$ is spontaneously broken by a scalar coupled to the Higgs sector of the theory, which will remain unspecified, and where the only new light degree of freedom at low energies is the ALP. Since the UV model is anomaly-free, so are all the couplings of the ALP in this scenario. Taking these points into account, we can construct a DFSZ-like EFT for the ALP, valid just below the EW breaking scale, as
\be
\sL_{DFSZ} = \frac{1}{2}\left(\partial_\mu a\partial^\mu a-m_a^2 aa\right)-i\displaystyle\sum_{f=u,d,e} \frac{m_f}{v_a}\chi_f a\bar{f}\gamma_5 f\label{eq:DFSZ},
\ee
where the sum is understood to run over the three fermion families\footnote{Notice that we are considering the case of massless neutrinos, which will therefore not couple to the ALP at tree level. Examples of axion models where a Seesaw is implemented can be found in Refs.~\cite{FileviezPerez:2019fku,CentellesChulia:2020bnf,DiLuzio:2020qio,Arias-Aragon:2020qip,Quevillon:2020aij,delaVega:2021ugs,Chen:2021haa,Arias-Aragon:2022ats}.}. Additionally, in the original DFSZ model, $\chi_f$ are entirely fixed by the ratio of the vacuum expectation values of the two Higgs doublets. However, we do not specify the UV completion here and consider therefore the possibility of all of them being independent, meaning that the theory at this point presents five free parameters, namely $v_a$, $m_a$ and the three $\chi_f$.

\begin{figure} [t] \centering \Large
\begin{subfigure}{
\begin{tikzpicture}
\begin{feynman}
\vertex (v1);
\vertex [left=of v1] (vi) {$a$};
\vertex [above right=of v1] (v2);
\vertex [below right=of v1] (v3);
\vertex [right=of v2] (vo1){$X$};
\vertex [right=of v3] (vo2){$X$};

\diagram* {
	(v1) -- [fermion] (v2) -- [fermion] (v3) -- [fermion] (v1),
	(v2) -- [boson] (vo1),
	(v3) -- [boson] (vo2),
	(vi) -- [scalar] (v1)
	};
\end{feynman}
\end{tikzpicture}
}
\end{subfigure}
\begin{subfigure}{
\begin{tikzpicture}
\begin{feynman}
\vertex (v1);
\vertex [left=of v1] (vi) {$a$};
\vertex [above right=of v1] (v2);
\vertex [below right=of v1] (v3);
\vertex [right=of v2] (vo1){$\bar{f}$};
\vertex [right=of v3] (vo2){$f$};

\diagram* {
	(vo1) -- [fermion] (v2) -- [fermion] (v3) -- [fermion] (vo2),
	(v1) -- [boson,edge label=$X$] (v2),
	(v3) -- [boson,edge label=$X$] (v1),
	(vi) -- [scalar] (v1)
	};
\end{feynman}
\end{tikzpicture}
}
\end{subfigure}

\caption{On the left: ALP coupling to gauge bosons in a DFSZ-like scenario. On the right: ALP coupling to SM fermions in a KSVZ-like scenario. In both cases, $X$ can be any pair of gauge bosons in the EW broken phase that respects $SU(3)_C\times U(1)_{EM}$.}
\label{fig:DFSZ-Coupling}
\end{figure}
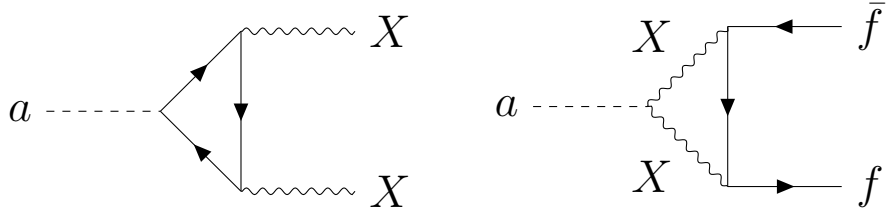

One point must be stressed at this stage. Because in the original DFSZ model the axion emerges only after the EW symmetry breaking (since part of it is made of $A$), it is not really adequate to consider a $SU(2)_L\otimes U(1)_Y$ invariant EFT like Eq.~\eqref{eq:ALPEFT}. Yet, it is always possible to switch to such a description by a chiral reparametrization of the fermion fields. The net effect is then to generate all the anomalous gauge couplings and derivative fermion couplings out of the pseudoscalar couplings in Eq.~\eqref{eq:DFSZ}. In that case, not only are the Wilson coefficients in Eq.~\eqref{eq:ALPEFT} correlated, but extreme care is needed when interpreting the EFT; in particular, the true gauge boson couplings are not directly those appearing in Eq.~\eqref{eq:ALPEFT}. For instance, Eq.~\eqref{eq:ALPEFT} seems to imply that $a \rightarrow \gamma\gamma, Z\gamma, ZZ, W^+W^-$ are correlated since they seem to depend only on two anomalous gauge couplings, but this is actually not true. The reason for that is the anomalies in the triangle graphs built on the derivative couplings of the ALP to fermions, see Fig.~\ref{fig:DFSZ-Coupling}. The Ward identities for both the vector and axial couplings are anomalous, and must be dealt with carefully. It is only once this is accounted for that the EFT in Eq.~\eqref{eq:DFSZ} can be matched onto an EFT of the form of Eq.~\eqref{eq:ALPEFT}, as shown in Ref.~\cite{Quevillon:2019zrd}\footnote{See also Refs.~\cite{Quevillon:2018mfl,Quevillon:2020hmx,Quevillon:2020aij,Bonnefoy:2020gyh,Quevillon:2021sfz,Filoche:2022dxl} for additional discussions on this matter.}.

In practice, it is much easier to simply use Eq.~\eqref{eq:DFSZ} as the starting EFT for a DFSZ-like ALP, and compute the gauge couplings via non-anomalous triangle graphs built on the pseudoscalar ALP couplings to fermions, Fig.~\ref{fig:DFSZ-Coupling}. Actually, this computation is well-known since it precisely corresponds to that of the usual 2HDM, up to a trivial rescaling of the final amplitudes, and the arbitrariness in the fermion couplings $\chi_{u,d,e}$. Though this is well-known, let us now provide explicit formulae for the coupling to each pair of gauge bosons. First, we define
\ba
\sL_{gauge}^{eff} = \frac{a}{4\pi v_a}\bigg(& g_{agg}{G^a}^{\mu\nu}\Tilde{G}^{a}_{\mu\nu}+g_{a\gamma\gamma}F_{\mu\nu}\Tilde{F}^{\mu\nu}\\
&+g_{aZ\gamma}Z_{\mu\nu}{\Tilde{F}}^{\mu\nu}+g_{aZZ}Z_{\mu\nu}\Tilde{Z}^{\mu\nu}+g_{aWW}{W^+}^{\mu\nu}{\Tilde{W}^{-}}_{\mu\nu}\bigg) \ ,
\label{eq:KSVZ_Broken}
\ea
where the effective couplings $g_{a V_1 V_2}$ are actually form-factors, i.e., functions of $m_{V_1}/m_a$ and $m_{V_2}/m_a$. Explicitly, these functions can be extracted from Ref.~\cite{Gunion:1991cw} as
\ba
g_{aV_{1}V_{2}} =-2i\pi \sigma%
\sum_{f=u,d,e}m_{f}\chi_{f}\left(  g_{V_1}^{f}g_{V_2}%
^{f^{\prime}}\mathcal{T}_{PVV}(m_{f})+g_{A_1}^{f}g_{A_2}%
^{f^{\prime}}\mathcal{T}_{PAA}(m_{f})\right)  \ .
\label{eq:DFSZgauge}
\ea 
In this expression, the fermion gauge couplings $\bar{\psi}_{f^{\prime}}(g_{V}^{f}\gamma^{\mu}-g_{A}^{f}\gamma^{\mu}\gamma_{5})\psi_{f}$ are
\begin{equation}%
\begin{tabular}
[c]{lllll}\hline
& $g$ & $W$ & $\gamma$ & $Z$\\\hline
$g_{V}^{f}\ =\ \ $ & $g_{s}T_{a}^{f}\ \ \ ,\ \ $ & $\dfrac{g}{\sqrt{2}}%
T_{3}^{f}\ \ ,\ \ $ & $eQ^{f}\ \ \ ,\ \ $ & $\dfrac{g}{2c_{W}}(T_{3}%
^{f}-2s_{W}^{2}Q^{f})\ ,$\\
$g_{A}^{f}\ =\ \ $ & $0\ \ \ ,\ \ $ & $\dfrac{g}{\sqrt{2}}T_{3}^{f}\ \ ,\ $ &
$0\ \ \ ,\ \ $ & $\dfrac{g}{2c_{W}}T_{3}^{f}\ ,$\\\hline
\end{tabular}
\end{equation}
where $SU(3)_C$ generators $T_{a}^{f}$ are normalized such that $\operatorname*{Tr}(T_{a}^{f}T_{b}^{f})=1/2\delta^{ab}$, $c_{W}=\cos\theta_{W},s_{W}=\sin\theta_{W}$, and CKM factors are understood. The parameter $\sigma$ accounts for the absence of crossed diagrams for $W^{+}W^{-}$, with $\sigma=1/2$ for $V_{1,2}=W$, $\sigma=1$ for $V_{1,2}=g,\gamma,Z$. The colour trace for quarks brings a further factor $1/2$ for $V_{1,2}=g$, and $N_{C}$ for $V_{1,2}=W,Z,\gamma$. Finally, the loop functions $\mathcal{T}_{PVV}^{\alpha\beta}(m_{f})$ and $\mathcal{T}_{AVV}^{\alpha\beta}(m_{f})$ are 
\ba
\mathcal{T}_{PVV}(m) & =\frac{-i}{2\pi^2}mC_{0}(m^{2})\ ,\\
\mathcal{T}_{PAA}(m) & =\frac{-i}{2\pi^2}m(C_{0}(m^{2})+2C_{1}(m^{2}))\ ,
\ea
where the standard three-point Passarino-Veltman scalar loop functions are $C_{i}(m_f^2)=C_{i}(m_{V_1}^{2},m_{V_2}^{2},m_{a}^{2},m_f^{2},m_f^{2},m_f^{2})$ for $V_{1,2}=g,\gamma,Z$, and  $C_{i}(m_f^2)=C_{i}(m_{W}^{2},m_{W}^{2},m_{a}^{2},m_{f}^{2},m_{f^{\prime}}^{2},m_{f}^{2})$, with $f^{\prime}$ the $SU(2)_L$ partner of $f$ for $W^{+}W^{-}$. Summation over the three families is understood in Eq.~\eqref{eq:DFSZgauge}, upon which the CKM factors disappear. Explicit expressions for the Passarino-Veltman scalar loop functions, as well as various limiting cases, can be found in Ref.~\cite{Gunion:1991cw}.

As stressed in Ref.~\cite{Quevillon:2019zrd}, the gauge couplings $g_{aV_{1}V_{2}}$ all vanish in the limit of massless fermions, as they should since in that case the ALP no longer couples to SM fields in the DFSZ scenario. Had we started from Eq.~\eqref{eq:ALPEFT}, such a vanishing would still occur of course, but require many delicate cancellations between the tree-level contributions from the gauge couplings, and loop-level, anomalous contributions from the derivative couplings. Evidently, such cancellations only occur once the proper correlations among the Wilson coefficients of Eq.~\eqref{eq:ALPEFT} are enforced. To be more specific and clearly illustrate the cancellations at play in Eq.~\eqref{eq:ALPEFT}, consider the $g_{aZZ}$ coupling. From the derivative interactions in Eq.~\eqref{eq:ALPEFT}, there are three different fermion loops corresponding to the axial-vector-vector,
vector-axial-vector, and axial-axial-axial triangles, with the first referring to the ALP coupling, and the last two to gauge couplings:%
\begin{align}
g_{aZZ}^{AVV}  & =-\pi\sum_{f=u,d,e,\nu}\chi_{A}^{f}g_{V,Z}^{f}g_{V,Z}%
^{f}\times i(q_{1}+q_{2})_{\gamma}\mathcal{T}_{AVV}^{\gamma}(f)\ ,\\
g_{aZZ}^{VAV}  & =-\pi\sum_{f=u,d,e,\nu}\chi_{V}^{f}(g_{V,Z}^{f}g_{A,Z}%
^{f}+g_{A,Z}^{f}g_{V,Z}^{f})\times(-i)(q_{1}+q_{2})_{\gamma}\mathcal{T}%
_{VAV}^{\gamma}\ ,\\
g_{aZZ}^{AAA}  & =-\pi\sum_{f=u,d,e,\nu}\chi_{A}^{f}g_{A,Z}^{f}g_{A,Z}%
^{f}\times i(q_{1}+q_{2})_{\gamma}\mathcal{T}_{AAA}^{\gamma}\ ,
\end{align}
where $\chi_{V}^{f}=\chi^{f_{R}}+\chi^{f_{L}}$ and $\chi_{A}^{f}=\chi^{f_{R}}-\chi^{f_{L}}$. These loop functions need not be computed, as their divergences can be worked out using the anomalous Ward identities,%
\begin{align}
i(q_{1}+q_{2})_{\alpha}\mathcal{T}_{AVV}^{\alpha\beta\gamma}\varepsilon
_{\beta}^{\ast}\varepsilon_{\gamma}^{\ast}  & =2im\mathcal{T}_{PVV}+\frac
{1}{2\pi^{2}}\varepsilon^{\beta\gamma\mu\nu}q_{1\mu}q_{2\nu}\varepsilon
_{\beta}^{\ast}\varepsilon_{\gamma}^{\ast}\ \label{eq:loop1} ,\\
i(q_{1}+q_{2})_{\alpha}\mathcal{T}_{VAV}^{\alpha\beta\gamma}\varepsilon
_{\beta}^{\ast}\varepsilon_{\gamma}^{\ast}  & =\frac{1}{2\pi^{2}}%
\varepsilon^{\beta\gamma\mu\nu}q_{1\mu}q_{2\nu}\varepsilon_{\beta}^{\ast
}\varepsilon_{\gamma}^{\ast}\ \label{eq:loop2},\\
i(q_{1}+q_{2})_{\alpha}\mathcal{T}_{AAA}^{\alpha\beta\gamma}\varepsilon
_{\beta}^{\ast}\varepsilon_{\gamma}^{\ast}  & =2im\mathcal{T}_{PAA}+\frac
{1}{2\pi^{2}}\varepsilon^{\beta\gamma\mu\nu}q_{1\mu}q_{2\nu}\varepsilon
_{\beta}^{\ast}\varepsilon_{\gamma}^{\ast}\ \label{eq:loop3}.
\end{align}
With this, one recovers the result of Eq.~\eqref{eq:DFSZgauge} from the $\mathcal{T}_{PVV}$ and $\mathcal{T}_{PAA}$ pieces. So, the axion-like DFSZ model requires all the anomalous terms in these loops to precisely cancel with the local gauge couplings of Eq.~\eqref{eq:ALPEFT}, i.e.,%
\ba\label{eq:DFSZ-Condition}
\frac{1}{4\pi}\sum_{f=u,d,e,\nu}\chi_{A}^{f}g_{V,Z}^{f}g_{V,Z}^{f}-2\chi
_{V}^{f}g_{V,Z}^{f}g_{A,Z}^{f}+\chi_{A}^{f}g_{A,Z}^{f}g_{A,Z}^{f}%
=-\alpha\left(  \mathcal{N}_{L}/t_{W}^{2}+t_{W}^{2}\mathcal{N}_{Y}\right)  \ .
\ea
Note that this equation only has a solution when $\chi_{A,V}^{f}$ are
$SU(2)_{L}\otimes U(1)_{Y}$ symmetric\footnote{As shown in Ref.~\cite{Quevillon:2019zrd}, these charges are ambiguous in the DFSZ model because they can all be shifted by a constant corresponding to either the conserved baryon or lepton number, but this is inessential here.}, with then quite naturally
$\mathcal{N}_{L}\sim\chi^{\ell_{L}}+3\chi^{q_{L}}$ and $\mathcal{N}_{Y}%
\sim-1/3\chi^{q_{L}}+4/3\chi^{u_{R}}+2/3\chi^{d_{R}}-\chi^{\ell_{L}}%
+2\chi^{e_{R}}$. By contrast, if one takes for example $\chi_{A}^{f}=\chi_{V}^{f}=\chi^{f}$, then we must also enforce $\mathcal{N}_{L}=0$ and $\mathcal{N}_{Y}\sim4/9\chi^{u}+1/9\chi^{d}+\chi^{e}$. Thus, it is clear that one can work in the basis of Eq.~\eqref{eq:ALPEFT}, but this is hardly a convenient choice when a DFSZ-like UV structure is to be imposed.

The fact that the naive $SU(2)_L \otimes U(1)_Y$ correlations among the electroweak gauge couplings are broken is most clearly visible in the limit of infinite fermion masses. Thanks to the overall $m_f$ factors, the $g_{aV_{1}V_{2}}$ become constant in that limit and we find:
\ba
g_{agg} & = \alpha_s\big(\chi_u + \chi_d\big)\\
g_{a\gamma\gamma} & = \frac{2\alpha}{9} \big(4N_C \chi_u + N_C \chi_d + 9 \chi_e\big)\\
g_{aZ\gamma} & = \frac{\alpha}{6 c_W s_W} \big(2 N_C \chi_u + N_C\chi_d + 3\chi_e\big) - t_W g_{a\gamma\gamma}\\
g_{aZZ} & = \frac{\alpha}{6 c_W^2 s_W^2}\big(N_C \chi_u + N_C\chi_d + \chi_e\big) - 2 t_W g_{a\gamma Z} - t_W^2 g_{a\gamma\gamma}\\
g_{aWW} & = \frac{\alpha}{12 s_W^2}\big(2 N_C \chi_u + 2 N_C\chi_d +3 \chi_e\big),
\ea
with $t_W=\tan\theta_W$. Even though all the couplings depend only on three parameters, it is not possible to put Eq.~\eqref{eq:KSVZ_Broken} back into an $SU(2)_L \otimes U(1)_Y$ invariant form. Yet, phenomenologically, it is in principle possible to reconstruct the pseudoscalar couplings of the ALP out of the individual measurements of these decay channels since they have no tree-level contribution in a DFSZ-like scenario.

Let us close this subsection with a phenomenologically oriented discussion. First of all, we must specify that, though five free parameters appear in the Lagrangian of Eq.~\eqref{eq:DFSZ}, only the combination $\chi_f/v_a$ enters the observables. Thus, the number of physical parameters is in fact four: the three $\chi_f/v_a$ ratios and $m_a$. This implies that measuring four of the ALP-gauge boson couplings should be enough to predict all of the remaining ones at tree level. In other words, if, for example, we were able to experimentally measure $g_{agg}$, $g_{a\gamma\gamma}$ and $g_{aZ\gamma}$ only one degree of freedom would remain. This implies that in the space spanned by $g_{aZZ}$ and $g_{aWW}$, the correlation would show up as a curve, as opposed to what would happen in the pure EFT point of view of Eq.~\eqref{eq:ALPEFT}, where the measurement of the same three ALP-gauge boson couplings would restrict the values of $g_{aZZ}$ and $g_{aWW}$ to a single point, as only two parameters, $\cN_L$ and $\cN_Y$, enter those couplings at tree level. If one were to include the SM fermion loop contribution arising from Eq.~\eqref{eq:ALPEFT}, which are indeed explicitly described by Eqs.~\eqref{eq:loop1} to \eqref{eq:loop3}, the parameters describing fermionic ALP couplings would enter the 1-loop ALP-gauge boson coupling, with the correlation no longer being a point, but being also different from the one observed in our approach, as the tree level ALP-gauge boson coupling still exists. If one were to further impose the DFSZ-like behaviour, i.e., Eq.~\eqref{eq:DFSZ-Condition}, even with the starting point of Eq.~\eqref{eq:ALPEFT}, we recover exactly the result displayed here.

\begin{figure}[t]
    \centering
    $g_{a\gamma\gamma}/v_a=0$\\ \vspace{5 pt}
    \includegraphics[height=0.3\textwidth,trim=1 1 32 1, clip]{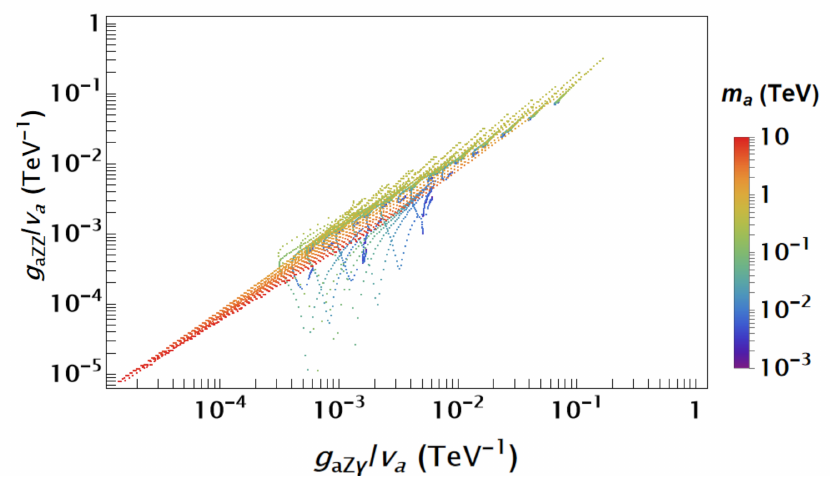}
    \includegraphics[height=0.3\textwidth,trim=3 1.5 1 1, clip]{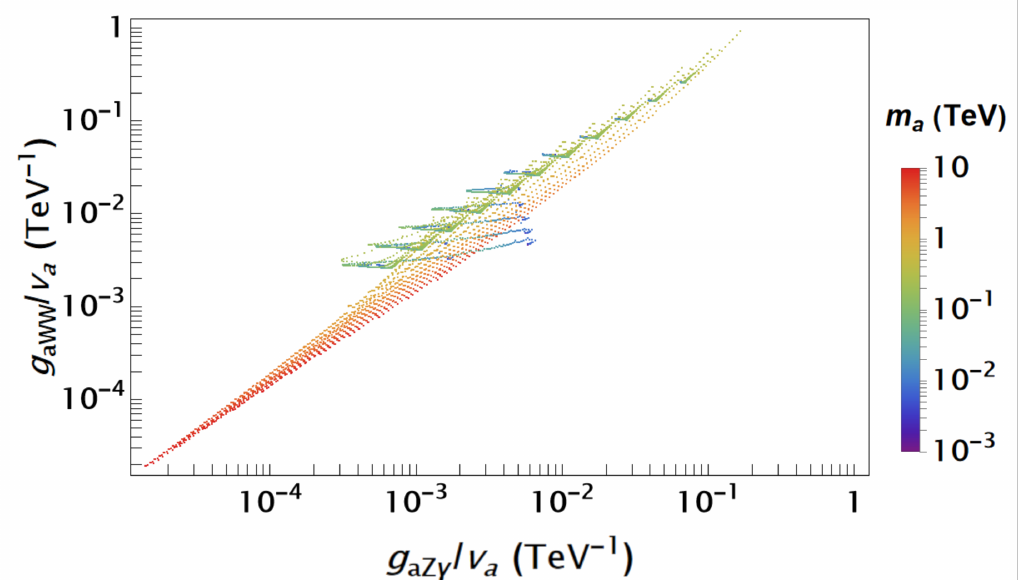}\\
    \caption{Prediction for the ALP-EW gauge boson coupling projected in the planes $(g_{aZZ},g_{aZ\gamma})$ and $(g_{aWW},g_{aZ\gamma})$ (from left to right) and their dependence on the ALP mass $m_a$ in the photophobic scenario for $v_a=1\TeV$. In this plot we scanned logarithmically all three parameters, $m_a$, $\chi_u$ and $\chi_d$, with both $\chi_u$ and $\chi_d$ varying between $0.1$ and $10$, and $m_a$ ranging between $1\GeV$ and $10\TeV$. Note that direct bounds on the ALP-fermion couplings have not been enforced, and may rule out part of these points.}
    \label{fig:DFSZ-Photophobic}
\end{figure}

Taking this freedom into account it is interesting to consider some particular situations. For example,  let us assume the (potentially fine-tuned) photophobic scenario in which $g_{a\gamma\gamma}=~0$. Such a constraint in the framework of Eq.~\eqref{eq:ALPEFT} would imply $\cN_Y=-\cN_L$, with only one remaining free parameter entering the ALP coupling to EW gauge bosons. In our sceneario, however, this immediately implies a relation between the four free parameters, which we can write as

\be
\chi_e=-\frac{\displaystyle\sum_{f=q} N_C Q_f^2 \chi_{f}\mathcal{A}_{1/2}^A\left(\frac{m_a^2}{4m_f^2}\right)}{\displaystyle\sum_{f=\ell} \mathcal{A}_{1/2}^A\left(\frac{m_a^2}{4m_f^2}\right)}\ ,
\ee
where the sum in the numerator runs over all quarks with flavour universal $\chi_u$ and $\chi_d$, and the one in the denominator runs over all three charged leptons, and the function $\mathcal{A}_{1/2}^A$ is defined in Appendix~\ref{App:DFSZ-Couplings}.

As previously discussed, such a situation where $g_{a\gamma\gamma}=0$ would imply $\cN_Y=-\cN_L$ in the naive EFT of Eq.~\eqref{eq:ALPEFT}, so the remaining pairs of ALP-EW gauge boson couplings would be proportional to each other, since they would depend linearly on a single free parameter. However, as we show in Fig.~\ref{fig:DFSZ-Photophobic}, the larger number of free parameters, three after fixing $g_{a\gamma\gamma}=0$, allows for more freedom for the different pairs of EW gauge boson couplings to the ALP, spanning areas, with a spread of not more than an order of magnitude if we avoid further fine-tuning.

It is particularly interesting to notice that even when the ALP-photon coupling exactly vanishes, it is possible to have sizeable couplings to all of the remaining EW gauge bosons, as well as possible hierarchies among them. Notice also that fixing $\chi_e$ in this scenario does not imply anything for $g_{agg}$, which remains with the same freedom as in the unconstrained scenario.

\subsection{A more constrained scenario}

Let us now consider a further reduction of the number of free parameters. There are indeed many phenomenologically motivated scenarios, like the photophobic one we just discussed, leptophobic ALPs with vanishing $\chi_e$, or leptophilic ones, where $\chi_u=\chi_d=0$.
In this section, however, we will consider an ALP that mimics the DFSZ axion couplings to fermions (see Refs.~\cite{Marsh:2015xka,DiLuzio:2020wdo} for reviews on axion models). In that model, two Higgs doublets are present; $H_u$ coupling to up quarks and $H_d$ coupling to down quarks and charged leptons (which would correspond to a 2HDM type-II). These two Higgs doublets couple to an additional scalar $\phi$ that has a VEV $v_\phi$ much larger than those of the Higgs doublets, $v_u$ and $v_d$. In terms of our Lagrangian from Eq.~\eqref{eq:DFSZ}, this implies that the formerly free couplings $\chi_f$ now take the following values:

\be
\chi_u=\frac{x^2}{1+x^2},\ \chi_d=\chi_e=\frac{1}{1+x^2},
\ee
where $x=\tan\beta=v_u/v_d$ is the ratio of the VEVs of the two Higgs doublets. In this setup, the number of free parameters has been reduced from four to three, namely $v_a$, $m_a$ and $x$. Notice that, while in the generic case, it is impossible to separately fix $\chi_f$ and $v_a$, since only the ratios $\chi_f/v_a$ enter observables, it is now possible to do so since specific values are prescribed for the $\chi_f$.
This means that if we were to measure experimentally two of the $\chi_f$ in this setup, we could immediately extract the values of $x$ and, more importantly, the ALP energy scale $v_a$.

\begin{figure}[t]
    \centering
    \includegraphics[height=0.27\textwidth,trim=1 2 30 1, clip]{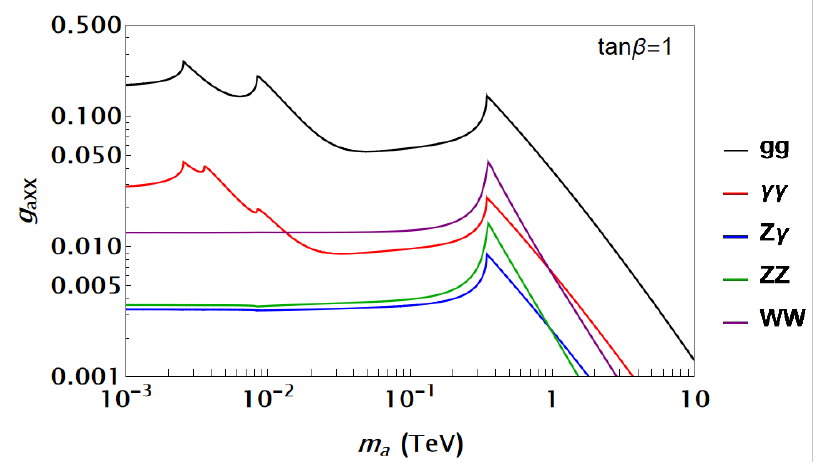}
    \includegraphics[height=0.27\textwidth,trim=13 2 2 1, clip]{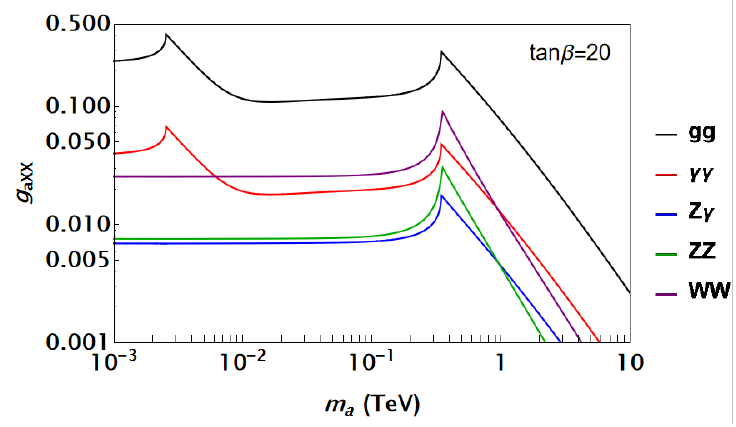}
    \caption{Predicted ALP couplings to gauge bosons with $x=1$ on the left and $x=20$ on the right. The different peaks correspond to kinematic thresholds, namely $m_a=2m_f$ for different fermions, which are almost lost in the presence of EW gauge bosons due to their high mass.}
    \label{fig:Democratic_DFSZ_Gauge}
\end{figure}

It is interesting to study the possible correlations among EW gauge boson couplings. Whereas in the usual EFT approach of Eq.~\eqref{eq:ALPEFT}, two free parameters enter linearly these couplings, namely $\cN_L$ and $\cN_Y$, and consequently, correlations are expected among the four ALP-EW gauge boson couplings, in the constrained scenario under consideration now we have shown that these couplings depend on the ALP mass $m_a$ and $x$ non-linearly. This implies that the correlation among EW couplings is highly non-trivial, unlike what one would expect from the EFT Lagrangian. In particular, as shown in Fig.~\ref{fig:Democratic_DFSZ_Gauge}, these couplings do depend on the ALP mass, whereas the expected behaviour from Eq.~\eqref{eq:ALPEFT} would be flat lines. Quite naturally, the predicted couplings present peaks when the ALP mass approaches $2m_f$ for different fermions; in particular, the charm quark, tau lepton, bottom quark and top quark mass thresholds can be clearly seen for $\tan\beta=1$. Increasing $\tan\beta$, however, makes the loop-induced ALP-gauge boson couplings be dominated by the contributions coming from up-type quarks, eliminating the peaks associated with the bottom quark and tau lepton. Notice also that the ALP coupling to photons and gluons have essentially the same dependence on the ALP mass for $\tan\beta=20$.

Finally, in this same constrained DFSZ scenario, one could try to illustrate how, after fixing $v_a$, the remaining two-dimensional parameter space formed by $\tan\beta$ and $m_a$ is constrained by the ALP-photon coupling. This coupling has been widely discussed in the literature, through searches for mono-photons with missing energy and tri-photon searches at LEP~\cite{Mimasu:2014nea}, as well as tri-photon searches at CDF and LHC~\cite{Jaeckel:2015jla,Knapen:2016moh}. Such bounds would hold up to roughly $v_a=100\GeV$, which implies an ALP scale below the EW one, whereas for higher ALP scales the parameter space of the ALP discussed here is unconstrained in what regards the ALP-photon coupling.

\begin{figure}[H]
    \centering
    \includegraphics[width=0.47\textwidth,trim=1 2 2 2, clip]{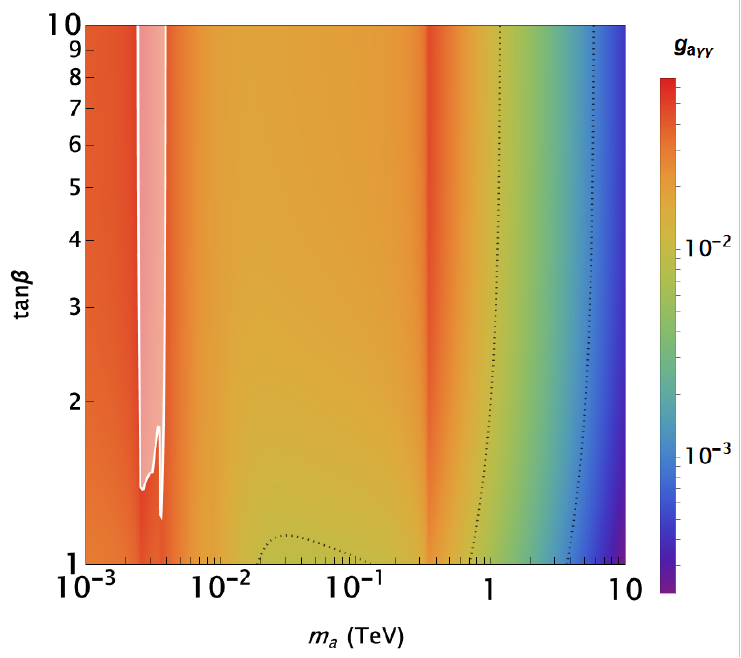}
    \caption{Parameter space for an ALP mimicking DFSZ axion couplings to fermions, with $v_a=100\TeV$. The colour code on the background shows how the induced ALP-photon coupling depends on the two free parameters, $\tan\beta$ and $m_a$; the DFSZ-like ALP couplings to massive gauge bosons present a different dependence. This is in contrast to the linear one on $\cN_Y$ and $\cN_L$ that all ALP-EW gauge boson coupling present in the pure EFT approach. A naïve projection of FCC-ee reach is shown in white.}
    \label{fig:Democratic_DFSZ_Bounds}
\end{figure}

However, we can look at prospects from future colliders, like the FCC, which will be able to probe even smaller ALP-photon couplings~\cite{EuropeanStrategyforParticlePhysicsPreparatoryGroup:2019qin,Bauer:2018uxu}, where at least 4 signal events are expected of an ALP decaying previous to the electromagnetic calorimeter. The prospects shown there (see Fig.~8.18 in Ref.~\cite{EuropeanStrategyforParticlePhysicsPreparatoryGroup:2019qin} and Fig.~6 inf Ref.~\cite{Bauer:2018uxu}), however, make some assumptions that are not fit for our approach. In particular, they assume a vanishing ALP-W boson coupling, $g_{aWW}=0$, as well as $\text{Br}(a\rightarrow\gamma\gamma)=1$, which are not necessarily true in our framework. Just for the sake of illustration, we show the parameter space span by $\tan\beta$ and $m_a$ for $v_a=100\TeV$ in Fig.~\ref{fig:Democratic_DFSZ_Bounds}, with the region probed by the FCC shaded in white.

We stress, however, that the approach of Ref.~\cite{Bauer:2018uxu} may cause the prospect presented there to not be applicable at face value in many situations, which is why we point at the need of more generic analyses. Notice as well that the absence of other bounds does not imply that the remaining parameter space is free: indeed, dedicated analyses considering the free parameters presented in this section should be performed to fully constrain the parameter space. This means considering bounds also on the ALP couplings to SM fermions, like those studied in Refs.~\cite{Bonilla:2021ufe,Bauer:2017ris,Bauer:2021mvw}, but using the expressions given in this paper for the ALP-gauge boson couplings. Such an analysis, though of the utmost interest for the community, remains beyond the scope of this letter, where our goal is to simply put forward the building blocks of a DFSZ-like ALP EFT.

%%%%%%%%%%%%%%%%%%%%%%%%%%%%%%%%%%%%%%%%%%%%%%%%%%%%%%%%%%%%
\section{KSVZ-like ALPs}\label{sec:KSVZ}

In this section, we proceed to carry a similar analysis to the one performed in the previous case but taking the KSVZ scenario as a guide. For the true QCD axion, the KSVZ scenario is built by introducing a new heavy coloured fermion, vector-like under the SM gauge symmetries, but not under $U(1)_{PQ}$. As none of the SM fields are charged under $\UPQ{}$, the low-energy axion effective theory obtained after integrating out the heavy fermion contains only anomalous $SU(2)_L \otimes U(1)_Y$ invariant couplings of the axion to the gauge fields. So, based on this, we construct a generic KSVZ-like ALP effective Lagrangian as
\ba
\sL_{KSVZ} &= \frac{1}{2}\left(\partial_\mu a\partial^\mu a-m_a^2 aa\right)\\
&+ \frac{a}{16\pi^2 v_a}\left(g_s^2\cN_C G_{\mu\nu}^a\Tilde{G}^{a,\mu\nu}+g^2\cN_L W_{\mu\nu}^i\Tilde{W}^{i,\mu\nu}+{g^\prime}^2\cN_Y B_{\mu\nu}\Tilde{B}^{\mu\nu}\right) \ ,
\label{eq:KSVZ}
\ea
without any of the derivative couplings of the axion to SM fermions. After EWSB, this Lagrangian can be projected onto the broken basis for the gauge bosons, giving back Eq.~\eqref{eq:KSVZ_Broken} with
\ba
g_{agg} & = \alpha_s \cN_C \ ,\\
g_{a\gamma\gamma} & = \alpha \left(\cN_L + \cN_Y\right) \ ,\\
g_{a\gamma Z} & = 2\alpha \left(-\cN_L/t_W + t_W\cN_Y\right) \ ,\\
g_{aZZ} & = \alpha \left(\cN_L/t_W^2 + t_W^2\cN_Y\right) \ ,\\
g_{aWW} & =\frac{2\alpha}{s_W^2} \cN_L \ .
\label{eq:KSVZ_Phys}
\ea
Contrary to the DFSZ-like ALP scenario, the ALP mass does not appear in these couplings as a free parameter, and the correlations are different, being linearly dependent on the free parameters $\cN_Y$ and $\cN_L$ in the KSVZ case.

Though the ALP does not couple at tree-level to the SM fermions, they are induced at one loop via the diagram shown on the right of Fig.~\ref{fig:DFSZ-Coupling}. Since the gauge couplings explicitly break the ALP shift symmetry, the induced ALP-fermion interaction should match onto

\be
\sL_{fermion}^{eff} = \sum_{f = u,d,e} \frac{m_f}{v_a} c_{af} a\bar{f}\gamma_5 f  \ .
\label{eq:KSVZ-ceff}
\ee
As before, summation over fermion flavors is understood. We can explicitly compute the different amplitudes to find the following expression for the coupling $c_{af}$:
\ba
c_{af} = & 16\Bigg(\alpha^2Q_f^2\left(\cN_{L}+\cN_{Y}\right)+\alpha_s^2\frac{4}{3}\cN_{C}\Bigg)I_0-\frac{\alpha^2(\cN_{L}/t_W^2+t_W^2\cN_{Y})}{s_W^2c_W^2}I_{ZZ}\\
&+\frac{16\alpha^2Q_f(T_f^3-2Q_fs_W^2)(-\cN_{L}/t_W+t_W\cN_{Y})}{s_Wc_W}I_{\gamma Z}-\frac{4\alpha^2\cN_L}{s_W^4}\displaystyle\sum_{f^\prime}V_{ff^\prime}I_{WW}
&,\label{eq:KSVZ-Coupling}
\ea
where $Q_f$ is the electric charge of the fermion $f$ under consideration, $V_{ff^\prime}$ is the fermion mixing matrix (CKM in the quark sector and identity in the case of massless neutrinos for leptons), while the scalar integrals are:
\ba
I_0 &= \displaystyle\int \frac{d^dk}{(2\pi)^d}\frac{k^2-(k\cdot p)^2/m_a^2}{k^2(p-k)^2\left((p_1-k)^2-m_f^2\right)} \ ,\\
I_{\gamma Z} & = \displaystyle\int \frac{d^dk}{(2\pi)^d}\frac{k^2-(k\cdot p)^2/m_a^2}{(k^2-M_Z^2)(p-k)^2\left((p_1-k)^2-m_f^2\right)} \ ,\\
I_{ZZ} &= \displaystyle\int \frac{d^dk}{(2\pi)^d}\frac{\left((k\cdot p)-2(k\cdot p_1)+2F_fk^2\right)-2F_f(k\cdot p)^2/m_a^2}{\left((p_1-k)^2-m_{f}^2\right)(k^2-M_Z^2)\left((p-k)^2-M_Z^2\right)} \ ,\\
I_{WW} &= \displaystyle\int \frac{d^dk}{(2\pi)^d}\frac{\left((k\cdot p)-2(k\cdot p_1)+2k^2\right)-2(k\cdot p)^2/m_a^2}{\left((p_1-k)^2-m_{f^\prime}^2\right)(k^2-M_W^2)\left((p-k)^2-M_W^2\right)} \ ,
\label{eq:KSVZ-LoopIntegral}
\ea
with $p$ and $p_1$ the ALP and fermion momenta respectively, $m_{f^\prime}$ the mass of the $SU(2)_L$ partner of the fermion $f$ and $F_f=1-4\left|Q_f\right|s_W^2+8\left|Q_f\right|^2s_W^4$. These integrals are written explicitly in terms of the Passarino-Veltman functions in App.~\ref{App:KSVZ-Couplings}. With these expressions we can illustrate the bounds on the ALP-photon coupling. In order to do so, we will consider a simplified scenario where $\cN_C=\cN_L=\cN_Y\equiv\cN_X$ and the only free parameters in the theory are the ratio $\cN_X/v_a$ and the ALP mass $m_a$.

\begin{figure}[H]
    \centering
    \includegraphics[height=0.435\textwidth,trim=1 2 2 1, clip]{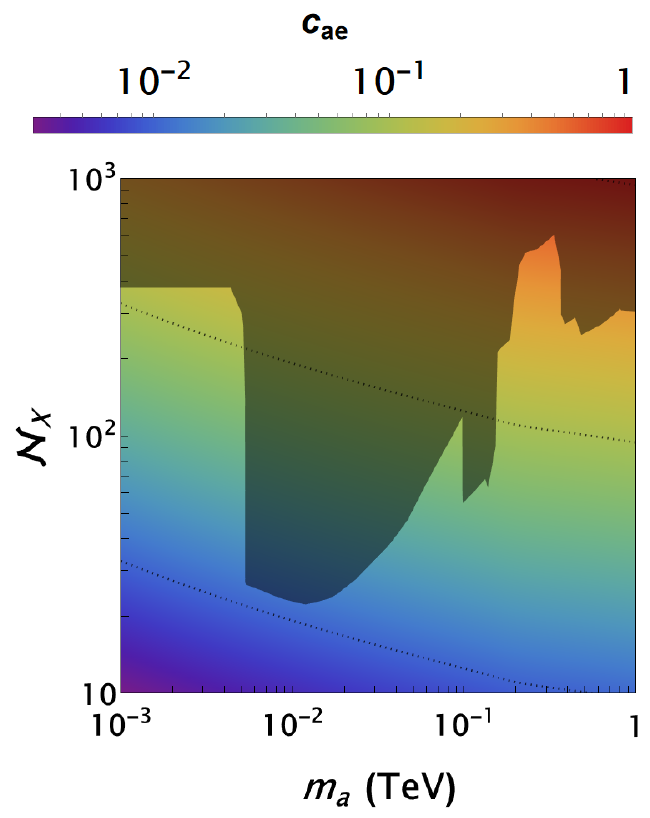}
    \includegraphics[height=0.435\textwidth,trim=14.5 2 2 1, clip]{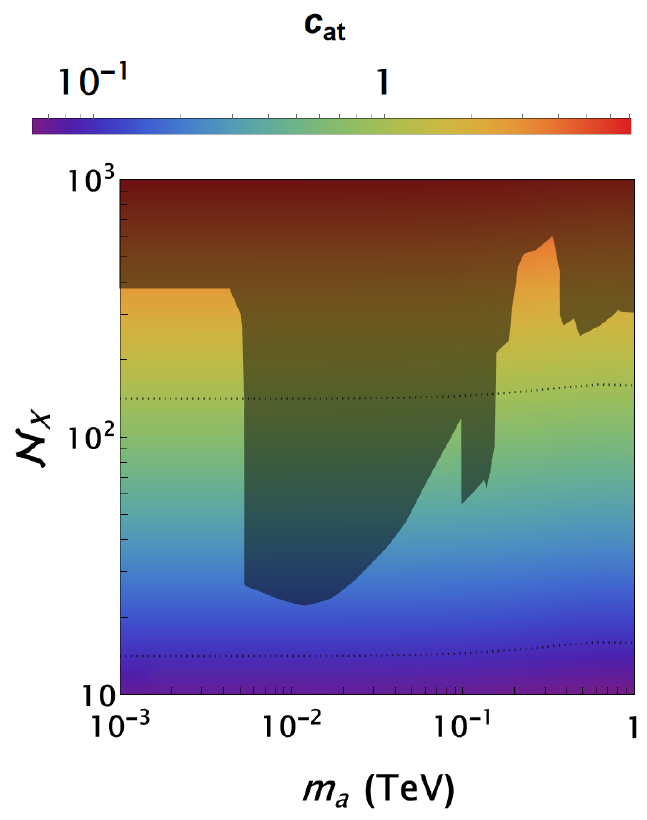}
    \includegraphics[height=0.435\textwidth,trim=14.5 2 1 1, clip]{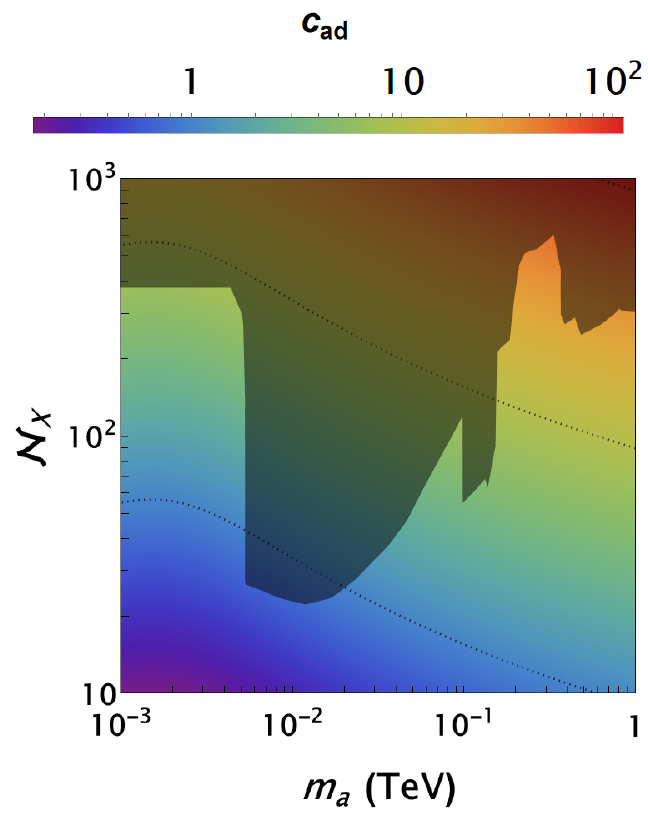}
    \caption{Parameter space for a KSVZ-like ALP coupling with the same strength to all gauge bosons in the unbroken phase for $v_a=1\TeV$. The colour code shows the induced fermion coupling, with the electron, top and down from left to right. The shadowed region is excluded due to limits on the ALP coupling to photons~\cite{Mimasu:2014nea,Jaeckel:2015jla,Knapen:2016moh}. The results in the $m_a\sim 1\TeV$ should be taken carefully due to the possible non-validity of the EFT. The loop integrals were evaluated numerically using \textit{LoopTools}~\cite{Hahn:1998yk,vanOldenborgh:1989wn}.}
    \label{fig:Democratic_KSVZ}
\end{figure}

Figure~\ref{fig:Democratic_KSVZ} shows how the parameter space for this simplified KSVZ scenario is constrained, with the colour code in the background representing now the induced ALP-fermion couplings, while the shaded dark region is obtained by limits on the ALP-photon coupling through searches for mono-photons with missing energy and tri-photon searches at LEP~\cite{Mimasu:2014nea}, as well as tri-photon searches at CDF and LHC~\cite{Jaeckel:2015jla,Knapen:2016moh}. As in the DFSZ case, we stress that this does not aim to be a full analysis on all the existing bounds, and further study would be required to include all bounds. This section, however, represents a guideline for dealing with a KSVZ-like ALP, with particular attention on the scalar integrals involved in the ALP-fermion couplings. These functions, however, are not UV finite, which raises two issues concerning renormalization scale and scheme, especially given the explicit presence of the four-dimensional Levi-Civita symbol in the gauge couplings. Let us now discuss these matters more in depth.

\subsection{On the $\gamma_5$ scheme}

Through the loop calculation, the Levi-Civita tensors in the gauge couplings of Eq.~\eqref{eq:KSVZ} must translate into the $\gamma_5$ couplings to the fermions in the effective Lagrangian Eq.~\eqref{eq:KSVZ-ceff}. Both these objects are intrinsically four dimensional, so there are ambiguities when dealing with them in dimensional regularization. Fixing these ambiguities as in Refs.~\cite{Bauer:2017ris,Bonilla:2021ufe} adds some scheme-dependence that, as we now show, can be alleviated.

The basic idea is to project from the start the amplitude into the right fermion state. Since the fermion pair must be in a pseudoscalar $J^{PC} = 0^{-+}$ state, this is done with~\cite{Martin:1970ai}
\be
P_{S=0}=\frac{1}{2\sqrt{2p^2}}\left(-\frac{1}{2}\varepsilon_{\rho\sigma\mu\nu}(p_1^\rho p_2^\sigma-p_1^\sigma p_2^\rho)\sigma^{\mu\nu}+(p^2-2m_f(\slashed{p}_1+\slashed{p}_2))\gamma_5  \right) \ ,
\label{eq:proj}
\ee
with $\sigma^{\mu\nu}=i(\gamma^\mu\gamma^\nu-\gamma^\nu\gamma^\mu)$ and $p$, $p_1$ and $p_2$ the ALP, fermion and anti-fermion momenta respectively. With this projector, we can extract:
\be
\mathcal{M}(a\rightarrow f \bar{f})=\bar{u}(p_1) T(a\rightarrow f \bar{f}) v(p_2)=\bar{u}(p_1) \gamma_5 v(p_2) F(a\rightarrow f \bar{f}),
\ee
\be
F(a\rightarrow f \bar{f})=\frac{1}{\sqrt{2p^2}}\Tr{(P_{S=0} T(a\rightarrow f \bar{f}))} \ .
\ee
This results in all the Dirac structure factoring out of the loop integral, in particular the $\gamma_5$, and the remaining scalar integral can be regularized without any ambiguities. Though in some sense, one could view this whole procedure as an alternative scheme to treat $\gamma_5$ in dimensional regularization, it is now based directly on the physical properties of the process and, in particular, on the conservation of energy-momentum and angular momentum. This same procedure is typically employed in dealing with the topologically similar $K_{L}\rightarrow \gamma\gamma \rightarrow \mu^+\mu^-$ amplitude, and permits meaningful matching with modelling of the short-distance dynamics hidden in the $K_{L}\rightarrow \gamma\gamma$ vertex (see for example Ref.~\cite{Isidori:2003ts}). 

Let us be a bit more quantitative and illustrate the property of the projector scheme in the case of the photon contributions in the triangle graph on the right of Fig.~\ref{fig:DFSZ-Coupling}. Looking only at the leading logarithm and finite constant term that arise from the divergent loop integrals, the scheme of Ref.~\cite{Bauer:2017ris,Bonilla:2021ufe}, where Larin's scheme~\cite{Larin:1993tq} is implemented, leads to
\be
c_{af}^{\gamma\gamma} = c \bigg(D_{\epsilon} + \ln\frac{\mu^2}{m_f^2} - \frac{4}{3} \bigg)+...\ ,
\label{eq:ccoeffN}
\ee
where $c=Q_f^2 \alpha^2 \left(\cN_L+\cN_Y\right)/(4\pi^2)$, $D_{\epsilon}$ is the usual $d-4$ pole and constants of the $\overline{MS}$ scheme, and the `$+...$' stands for the functional dependencies on the masses. The projector scheme rather leads to \cite{Isidori:2003ts}
\be
c_{af}^{\gamma\gamma} = c \bigg(D_{\epsilon} + \ln\frac{\mu^2}{m_f^2} + \frac{5}{3} \bigg)+...\ .
\label{eq:ccoeffT}
\ee
The prefactors, logarithms and remainders (the `$+...$') match, but not the finite terms. Though similar in magnitude, they appear with opposite signs, and induce a different asymptotic value for the $c_{af}$ coefficient in the limit $m_a \ll m_f$. As said before, we think the projector provides a more physically meaningful regularization scheme and should be preferred. To confirm this though, it is necessary to go beyond the triangle graph and consider the matching with the full theory. This is done in the next section. 

\subsection{On the renormalization scale}

If one considers the Lagrangian in Eq.~\eqref{eq:KSVZ} as the EFT valid at the electroweak scale, the form of Eq.~\eqref{eq:ccoeffT} would suggest first to set the renormalization scale $\mu$ to $M_W$, and add the fermionic effective operators of Eq.~\eqref{eq:KSVZ-ceff} with bare parameters $c_{af}^0$ to those in Eq.~\eqref{eq:KSVZ} (or more precisely, to the EFT just below the EW scale, once the EW symmetry is broken). Those would absorb the $D_{\epsilon}$ divergences of Eq.~\eqref{eq:ccoeffT}, but would add to the final physical $c_{af}$ some unknown $c_{af}^{ren}\sim\mathcal{O}(1)$ contributions. Though this is perfectly sound and follows standard EFT procedures, it goes contrary to the KSVZ-like assumption. In other words, a truly KSVZ-like scenario must not lead to $c_{af}$ coefficients of $\mathcal{O}(1)$ at the electroweak scale.

\begin{figure} [t] \centering \Large
\begin{subfigure}{
\begin{tikzpicture}
\begin{feynman}
\vertex (v1);
\vertex [left=of v1] (vi) {$a$};
\vertex [above right=of v1] (v2);
\vertex [below right=of v1] (v3);
\vertex [right=of v2] (vo1);
\vertex [right=of v3] (vo2);
\vertex [right=of vo1] (vof1){$f$};
\vertex [right=of vo2] (vof2){$\bar{f}$};

\diagram* {
	(v1) -- [fermion] (v2) -- [fermion, edge label'=$Q$] (v3) -- [fermion] (v1),
	(v2) -- [boson] (vo1),
	(v3) -- [boson] (vo2),
	(vof2) -- [fermion] (vo2) -- [fermion] (vo1) -- [fermion] (vof1),	
	(vi) -- [scalar] (v1)
	};
\end{feynman}
\end{tikzpicture}
}
\end{subfigure}
\caption{ALP coupling to SM fermions in a KSVZ-like scenario, where $Q$ are the heavy fermions carrying PQ charge.}
\label{fig:Two-Loop-Process}
\end{figure}
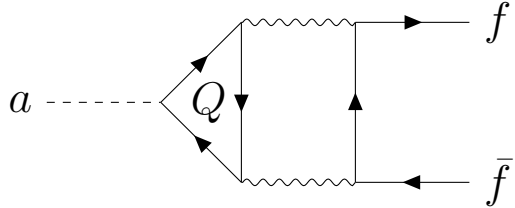

In a true KSVZ-like scenario, the fermionic couplings do arise, but from finite two-loop processes, through a loop of heavy quarks $Q$ (see Fig.~\ref{fig:Two-Loop-Process}). To explore the consequences, in terms of constraints on the one-loop results of Eq.~\eqref{eq:KSVZ-ceff}, let us first consider a simpler situation in which $a \rightarrow \bar{f}f$ arises only from two-loop processes with intermediate photons. That amplitude has been calculated long ago in the context of neutral meson decays to lepton pairs, and can be found in Ref.~\cite{Ametller:1983ec} (see also Ref.~\cite{Ecker:1991ru}). In the limit of vanishing ALP mass, it has the expansion
\be
c_{af}^{\gamma\gamma} = c \bigg(0 + \ln\frac{m_Q^2}{m_f^2} + \frac{17}{6} + \frac{5}{27}\frac{m_f^2}{m_Q^2}\ln\frac{m_Q^2}{m_f^2} +\frac{11}{54}\frac{m_f^2}{m_Q^2}+ ...\bigg)\ .
\label{eq:ccoeff2}
\ee
The first three terms are to be compared to Eq.~\eqref{eq:ccoeffN} and  Eq.~\eqref{eq:ccoeffT}. Obviously, there is no divergence here, while the scale is naturally set by the heavy quark mass, $\mu = m_Q$. The true finite term ends up being very close to that of Eq.~\eqref{eq:ccoeffT}. Note well that here, there is no scheme dependence of any sort since the amplitude is finite. This, in our opinion, justifies our earlier claim that using the projector Eq.~\eqref{eq:proj} allows for a physically more meaningful treatment of $\gamma_5$.

The correspondence with the EFT approach would suggest to split the logarithm in Eq.~\eqref{eq:ccoeff2} at the $M_W$ scale and define the renormalized $c_{af}^{ren}$ to appear just below the electroweak scale as $c_{af}^{ren} = c\ln(m_Q^2/M_W^2)$. Given that $c$ includes all the typical two-loop suppression factors, $c_{af}^{ren}$ is clearly not of $\mathcal{O}(1)$, even for $m_Q \gg M_W$. Though this is illustrated with the $\gamma\gamma$ intermediate state, the other cases should exhibit the same behaviors, there should be no significant tree-level contributions to the ALP-fermion couplings in a KSVZ-like ALP scenario. In practice, since we do not have the full two-loop result for intermediate weak bosons, we will use the one-loop results of Eq.~\eqref{eq:KSVZ-LoopIntegral}, set the scale at $\mu = v_a \approx m_Q$ considering how the heavy quarks receive their mass, and rely on the projector Eq.~\eqref{eq:proj} to give reasonable finite terms.

To close this section, let us give a different take on why one should set $\mu = v_a \approx m_Q$ in the one-loop results. This is clearly different from what one finds in the literature, seems to go against the EFT paradigm, and may be a bit puzzling given that SM fermions are massless above the EW scale. Naively, anything happening above $M_W$ looks irrelevant since the two-loop process is proportional to the fermion mass. There are two ways of understanding this. First, notice that in the KSVZ scenario, the electroweak and PQ symmetry breaking mechanisms are totally separated. So, in principle, the two-loop calculation can be done in the EW broken phase, there is no need to first integrate out the heavy fermions. The leading logarithm $\ln(m_Q^2/m_f^2)$ then arises naturally. 

A second way of understanding this, in line with the EFT picture, is to remember that not all effective operators are included in the Lagrangian Eq.~\eqref{eq:KSVZ}. If we were to precisely match a KSVZ scenario onto operators at the EW scale, we should add to Eq.~\eqref{eq:KSVZ} the dimension-five shift-symmetry-breaking (but $SU(2)_L \otimes U(1)_Y$ invariant) operators
\be
\sL_{KSVZ} \rightarrow \sL_{KSVZ} +  \frac{a}{v_a}  \big( c_{au} \bar{Q}_L \text{Y}_u u_R H +c_{ad} \bar{Q}_L \text{Y}_d d_R H^*+c_{ae} \bar{L}_L \text{Y}_e e_R H^{*}\big) \ .
\label{eq:KSVZ-cHeff}
\ee
The diagrams contributing to these operators are simply obtained from those in Fig.~\ref{fig:Two-Loop-Process} by adding a Higgs tadpole on the final fermion propagator (it cannot be attached anywhere else since the amplitude vanishes for $m_f=0$). These processes are finite, and adding these operators would simply induce the small $c_{af}^{ren} = c\ln(m_Q^2/M_W^2) \ll \mathcal{O}(1)$ corrections discussed before. So, an exact calculation from $m_Q$ down to the $M_W$ scale in the unbroken phase, followed by a calculation from $M_W$ down to $m_f$ in the broken phase, would not change our results.

%%%%%%%%%%%%%%%%%%%%%%%%%%%%%%%%%%%%%%%%%%%%%%%%%%%%%%%%%%%%

\section{Conclusions}\label{sec:Conclusions}

In this letter we have detailed the computation of loop generated ALP couplings in two possible scenarios. In a DFSZ-like ALP scenario, the SM fermions couple to the ALP at tree level, and the gauge bosons coupling arise at the one-loop level. Conversely, in a KSVZ-like ALP scenario, the SM gauge bosons are coupled to the ALP at tree level, and the SM fermion coupling to the ALP arise at the one-loop level. These basic patterns, inspired by those arising for the respective QCD axion models, are then quite constrained, and permit to identify simplified benchmarks for ALP searches.\\

In a DFSZ-like ALP model, the computation has been carried out in the non-derivative basis for the ALP-fermion couplings, to avoid consistency issues when computing loop couplings to gauge bosons. With this basis being manifestly renormalizable, neither divergences nor anomalies show up, and the calculation is completely analogous to the one performed in a 2HDM for the pseudoscalar Higgs decay to gauge bosons. The ALP-fermionic interactions introduce new parameters ($\chi_f$ in our notation) which on top of the ALP mass $m_a$ and decay constant $v_a$, allow to entirely describe the ALP-gauge sector, independently of the ALP-scalar sector. As it is known from Ref.~\cite{Quevillon:2019zrd}, the physical DFSZ-like ALP-gauge couplings do have an ALP-mass dependence unlike what is often inferred from usual generic ALP EFT (or simplified models).
When the ALP-fermion interactions strictly follow the type II 2HDM pattern, the resulting DFSZ-like ALP sector can be fully described by three parameters ($m_a$,$v_a$,$\tan\beta$). Again, the ALP-gauge sector is fully described by this simple parameter space, and ALP-gauge couplings do depend on the physical ALP mass. We gave an example on how one could easily use this parametrization to set bounds on DFSZ-like ALPs, considering only constraints on the ALP-photon coupling, which however do not set bounds on the constrained scenario presented when $v_a\gtrsim 100\GeV$. This, together with the naïve prospect of FCC-ee reach shown here, reinforces the need for a full analysis of current and future experimental limits for this framework.\\

In a KSVZ-like ALP model, the axion-gauge sector is parametrized by three parameters, (${\cal N}_C$, ${\cal N}_Y$, ${\cal N}_L$) on top of the ALP mass $m_a$ and decay constant $v_a$. These couplings imply a linear correlation among the gauge couplings in the EW broken phase and, unlike the DFSZ-like ALP scenario, the ALP mass does not appear in them. Regarding the ALP-fermionic sector, a KSVZ-like ALP does not couple at tree-level to the SM fermions, but these couplings are induced at one loop. There are two issues concerning these ALP-fermion couplings to consider. 

First, the ALP-fermion couplings generated at one-loop by the ALP gauge couplings are divergent. Thus, bare ALP-fermion couplings have to be added to the EFT to absorb these divergences. This adds some unknown $\mathcal{O}(1)$ contribution to the final physical ALP-fermion couplings. Though this is perfectly sound and follows standard EFT procedures, it goes contrary to the KSVZ-like assumption. A truly KSVZ-like scenario must not lead to sizable ALP-fermion couplings of $\mathcal{O}(1)$ at the electroweak scale, because such couplings can only arise at two loops in a full KSVZ-like theory. Actually, the renormalized coefficients which appear just below the electroweak scale should better be identified with the logarithm of the ratio of the two scales at play, $\ln(v_a^2/M_W^2)$, times some loop suppression factors. We have checked this explicitly for the ALP-two-photon coupling, for which the two-loop amplitude is known, and found that the net effect is simply to push the renormalization scale from $\mu = M_W$ to $v_a$ in the one-loop effective calculation. After taking into account the loop suppression factors, this results in a change of the ALP-fermion couplings far smaller than $\mathcal{O}(1)$, even for $v_a \gg M_W$. For the intermediate weak gauge boson contributions to the ALP-fermion couplings, the full two-loop amplitudes are not known but we expect them to behave similarly. So, we enforced our prescription and simply set $\mu = v_a$ in the usual one-loop EFT results.

A second issue is related to the Levi-Civita tensors present in the ALP-gauge boson couplings. Since the gauge-boson loops are divergent in the EFT, some scheme dependence enters when these tensors are interpreted in dimensional regularization. A convenient way to alleviate this scheme dependence is to project the amplitude onto the $0^{-+}$ fermion state. In this way, the $\gamma_5$ of the ALP-fermion couplings is factored out from the start, all loop integrals are scalar, and they give finite terms in good agreement with the full two-loop computation (which is finite, hence free of such scheme dependence). 

Using this procedure, together with the prescription for the renormalization scale discussed above, the ALP-fermionic sector is fully described by the five dimensional parameter space ($m_a$, $v_a$, ${\cal N}_C$, ${\cal N}_Y$, ${\cal N}_L$), and no new free parameters enter. For illustration, we have recasted the bounds on the ALP-photon coupling in this simplified parameter space, taking all tree-level ALP-gauge boson couplings to be the same.

We stress that the goal of this letter is to provide ALP parametrizations inspired from the two most common QCD axion models, DFSZ and KSVZ. Different correlations are present in each case, paving the way for a comprehensive and complete analysis. It is not our goal here to perform such a dedicated study, which includes all available bounds on ALP couplings and recasts as well the existing searches for 2HDM into the DFSZ-like ALP parameter spaces. Such an endeavour, together with the study of observables that may include ALP-Higgs couplings, is left for a future work. We believe also that the proposed ALP-benchmarks should be of interest for the experimental community when displaying their results for pseudo-scalar searches as they offer a comprehensive way to probe the truly axion-like ALP parameter space.

%%%%%%%%%%%%%%%%%%%%%%%%%%%%%%%%%%%%%%%%%%%%%%%%%%%%%%%%%%%%
\section*{Acknowledgements}
The authors acknowledge funding from the French Programme d'investissements d'avenir through the Enigmass Labex and support from the IN2P3 Master project ``Axions from Particle Physics to Cosmology''. J.Q. also aknowledges the support from the IN2P3 Master project BSMGA and the support from the ``Evenements rares - 2022'' call by CNRS. F.A.A. also acknowledges the support of the European Consortium for Astroparticle Theory in the form of an Exchange Travel Grant.

%%%%%%%%%%%%%%%%%%%%%%%%%%%%%%%%%%%%%%%%%%%%%%%%%%%%%%%%%%%%
\appendix
\section{DFSZ-like ALP couplings to gauge bosons}\label{App:DFSZ-Couplings}

The $g_{aV_1 V_2}$ effective couplings to gauge bosons appearing in Eq.~\eqref{eq:KSVZ_Broken} have the general expression in Eq.~\eqref{eq:DFSZgauge}.  For ease of use, in this appendix, explicit expressions for these loop functions are collected for each final state. Note that all of these expressions can be found or deduced from the literature, in particular from Ref.~\cite{Gunion:1991cw}.

\subsection*{Coupling to photons}

From Eq.~(2.24) in Ref.~\cite{Djouadi:2005gj}, the pseudoscalar Higgs decay width to two photons in a 2HDM reads
\be
\Gamma(a\xrightarrow[]{}\gamma\gamma)=\frac{\alpha^2m^3_a}{2^8\pi^3v^2}\left|\displaystyle\sum_f N_C Q_f^2 g_{aff}\mathcal{A}_{1/2}^A\left(\frac{m_a^2}{4m_f^2}\right)\right|^2,
\ee
where $N_C$ and $Q_f$ are the number of colours and electric charge of the fermion being considered, $g_{aff}=(v/v_a)\chi_f$, and the amplitude $\mathcal{A}_{1/2}^A$ is defined as follows:
\be
\mathcal{A}_{1/2}^A(x)=2x^{-1}f(x),\quad f(x)=\begin{cases} \arcsin^2{\sqrt{x}} & x\leq 1 \ ,\\ -\frac{1}{4}\left(\log{\frac{1+\sqrt{1-x^{-1}}}{1-\sqrt{1-x^{-1}}}}-i\pi\right)^2 & x>1 \ .\end{cases}
\label{eq:FormFactorA}
\ee
By comparing this expression with our effective coupling, we find
\be
g_{a\gamma\gamma}=\alpha\left| \displaystyle\sum_f N_C Q_f^2 \chi_{f}\mathcal{A}_{1/2}^A\left(\frac{m_a^2}{4m_f^2}\right)\right|.
\ee

\subsection*{Coupling to $Z\gamma$}

From Ref.~\cite{Gunion:1991cw} we know the expression for the decay width of a 2HDM pseudoscalar into a photon and a $Z$ boson:
\be
\Gamma(a\xrightarrow[]{}Z\gamma)=\frac{\alpha^2 m^3_a}{128\pi^3v^2s_W^2}\left(1-\frac{M_Z^2}{m_a^2}\right)^3\left|\displaystyle\sum_f \frac{\hat{v}_f}{c_W} N_C Q_fg_{aff} \mathcal{C}\left(M_Z^2,0;m_f^2\right)\right|^2,
\ee
where $\hat{v}_f=2T^3_f-4Q_f s_W^2$, with $T^3_f$ the weak isospin of the fermion considered and $s_W$ and $c_W$ being the sine and cosine respectively of the weak mixing angle. The form factor in this case reads:

\be
\mathcal{C}\left(M_Z^2,0;m_f^2\right) = \frac{2m_f^2}{m_a^2-M_Z^2}\left(f\left(\frac{m_a^2}{4m_f^2}\right)-f\left(\frac{M_Z^2}{4m_f^2}\right)\right),
\ee

Again, by comparing the explicit expression of the decay width with that we would find using our effective coupling we identify $g_{aZ\gamma}$ to be
\be
g_{aZ\gamma}=\frac{2\alpha}{s_W}\left|\displaystyle\sum_f \frac{\hat{v}_f}{c_W}N_C Q_f \chi_f \frac{m_f^2}{m_a^2-M_Z^2}\left(f\left(\frac{m_a^2}{4m_f^2}\right)-f\left(\frac{M_Z^2}{4m_f^2}\right)\right) \right|.
\ee

\subsection*{Coupling to $W^+W^-$}

The expression for the decay width of Ref.~\cite{Gunion:1991cw} is:
\bg
\Gamma(a\xrightarrow[]{}W^+W^-)=\frac{\alpha^2m_a^3}{2^9\pi^3s_W^4v^2}\left(1-\frac{4M_W^2}{m_a^2}\right)^{3/2}\left|A_f\right|^2,\\
A_f=\displaystyle\sum_f m_f^2 N_C g_{aff} \left(\mathcal{C}\left(m_a^2;M_W^2;m_f^2,m^2_{\tilde{f}}\right)+\mathcal{J}\left(m_a^2;M_W^2;m_f^2,m^2_{\tilde{f}}\right)\right).
\eg
The functions $\mathcal{C}$ and $\mathcal{J}$ are defined as the integrals:
\bg
\mathcal{C}\left(m_a^2;M_W^2;m_f^2,m^2_{\tilde{f}}\right) = \displaystyle\int^1_0 dx \displaystyle\int_0^x dy \frac{1}{D},\\
\mathcal{J}\left(m_a^2;M_W^2;m_f^2,m^2_{\tilde{f}}\right) = -\displaystyle\int^1_0 dx \displaystyle\int_0^x dy \frac{y}{D},
\eg
where the function $D$ reads
\be
D = m_a^2(1-x)(x-y)+M_V^2y(1-y)-m_f^2(1-y)-m^2_{\tilde{f}}y.
\ee
In all the previous formulae $f$ refers to the fermion to which the ALP couples, whereas $\tilde{f}$ corresponds to its $SU(2)$ partner, while $M_V$ corresponds to the mass of the vector boson involved in the process.

From this, we can identify our effective coupling to the $W$ bosons as:
\be
g_{aWW}=\frac{\alpha}{2s_W^2}\left|\displaystyle\sum_f m_f^2 N_C \chi_f  \left(\mathcal{C}\left(m_a^2;M_W^2;m_f^2,m^2_{\tilde{f}}\right)+\mathcal{J}\left(m_a^2;M_W^2;m_f^2,m^2_{\tilde{f}}\right)\right)\right|
\ee

\subsection*{Coupling to $ZZ$}

The explicit expression for the pesudoscalar decay to two $Z$ bosons is~\cite{Gunion:1991cw}:
\ba
\Gamma(a\xrightarrow[]{}ZZ)&=\frac{\alpha^2m_a^3}{2^{10}\pi^3v^2s_W^4c_W^4 }\left(1-\frac{4M_Z^2}{m_a^2}\right)^{3/2}\\
&\times\left|N_Cm_f^2g_{aff}\left(F_f\mathcal{C}\left(m_a^2;M_Z^2;m_f^2\right)+\mathcal{J}\left(m_a^2;M_Z^2;m_f^2\right)\right)\right|^2.
\ea
From here, we thus identify:
\be
g_{aZZ}= \frac{\alpha}{2s_W^2c_W^2}\left|\displaystyle\sum_f m_f^2 N_C \chi_f  \left(F_f\mathcal{C}\left(m_a^2;M_Z^2;m_f^2\right)+\mathcal{J}\left(m_a^2;M_Z^2;m_f^2\right)\right)\right|.
\ee

\subsection*{Coupling to gluons}

From Ref.~\cite{Djouadi:2005gj}, the pseudoscalar decay width to two gluons is:
\be
\Gamma(a\xrightarrow[]{}gg)=\frac{\alpha_s^2m^3_a}{72\pi^3v^2}\left|\displaystyle\sum_Q \frac{3}{4} g_{aQQ}\mathcal{A}_{1/2}^A\left(\frac{m_a^2}{4m_f^2}\right)\right|^2,
\ee
where the function $\mathcal{A}_{1/2}^A$ is defined in Eq.~\eqref{eq:FormFactorA}. Thus, we can identify the coupling constant as:
\be
g_{agg}=\frac{2\alpha_s}{3}\left| \displaystyle\sum_Q \frac{3}{4} \chi_{Q}\mathcal{A}_{1/2}^A\left(\frac{m_a^2}{4m_Q^2}\right)\right|,
\ee

\section{Loop integrals for the KSVZ-like ALP}\label{App:KSVZ-Couplings}

In this appendix we present the scalar loop integrals, shown in Eq.~\eqref{eq:KSVZ-LoopIntegral}, that enter the ALP couplings to gauge bosons in the KSVZ-like framework. Their expression in terms of Passarino-Veltman functions allows for their easy evaluation using numerical packages like \it{Looptools}~\rm\cite{Hahn:1998yk,vanOldenborgh:1989wn} as we did in this paper.

\be
I_0=\frac{i\pi^2}{4m_f^2}\left(3A_0(m_f^2)+m_f^2\left(2-m_a^2C_0(m_a^2,m_f^2,m_f^2,0,0,m_f^2)\right)\right),
\ee
\ba
I_{\gamma Z}&=\frac{i\pi^2}{8m_a^2m_f^2}\bigg(2m_f^2(m_a^2-M_Z^2)+2(2m_a^2-M_Z^2)A_0(m_f^2)-m_a^2A_0(M_Z^2)\\
&+(2m_a^2m_f^2+(m_a^2+2m_f^2)M_Z^2)B_0(m_f^2,m_f^2,M_Z^2)\\
&-2m_f^2(m_a^2-M_Z^2)^2C_0\left(m_a^2,m_f^2,m_f^2,0,M_Z^2,m_f^2\right)\bigg),
\ea
\ba
I_{ZZ}&=\frac{i\pi^2}{2m_f^2}\bigg(2m_f^2\left(B_0(m_a^2,M_Z^2,M_Z^2)-B_0(m_f^2,m_f^2,M_Z^2)\right)\\
&+F_f (A_0(m_f^2)-A_0(M_Z^2)+(2m_f^2+M_Z^2)B_0(m_f^2,m_f^2,M_Z^2))\\
&+m_f^2\left((1-F_f)m_a^2-2(1-2F_f)M_Z^2\right)C_0\left(m_a^2,m_f^2,m_f^2,M_Z^2,M_Z^2,m_f^2\right)\bigg),
\ea
\ba
I_{WW}&=\frac{i\pi^2}{2m_f^2}\bigg(A_0(m_{f^\prime}^2)-A_0(M_W^2)+2m_f^2B_0(m_a^2,M_W^2,M_W^2)\\
&+\left(m_f^2-m_{f^\prime}^2+M_W^2\right)B_0(m_f^2,m_{f^\prime}^2,M_W^2)\\
&+2m_f^2\left(m_{f^\prime}^2-m_f^2+M_W^2\right)C_0\left(m_a^2,m_f^2,m_f^2,M_W^2,M_W^2,m_{f^\prime}^2\right)\bigg).
\ea
%%%%%
%%%%%%%%%%%%%%%%%%%%%%%%%  Bibliography    %%%%%%%%%%%%%%%%%%%%%%%%
%%%%%

\footnotesize

\bibliography{biblio}{}
\bibliographystyle{BiblioStyle}

\end{document}